\documentclass[12pt]{article}

\usepackage{rotating} % Rotating table
\usepackage{graphicx}
\usepackage{amssymb}
\begin{document}

%\begin{frontmatter}

\title{Basic principles and concept design of a real-time clinical decision support system for managing medical emergencies on missions to Mars}

\author{Juan M Garcia-Gomez\\ {\small Universitat Politècnica de València,}\\ {\small Biomedical Data Science Lab,}\\ {\small Instituto Universitario de Tecnologías de la Información y Comunicaciones (ITACA),}\\ {\small Valencia, Spain}}

\maketitle

\begin{abstract}
	
	Space agencies and private companies prepare the beginning of human space exploration for the 2030s with missions to put the first human on the Mars surface.
	The absence of gravity and radiation, along with distance, isolation and hostile environments, are expected to increase medical events where previously unseen manifestations may arise.
	Therefore, crewmembers may have variate medical emergencies during the two to three years of spaceflight.
	The current healthcare strategy based on telemedicine and the possibility to stabilize and transport the injured crewmember to a terrestrial definitive medical facility is not applicable in exploration class missions.
	Therefore, the need for deploying the full autonomous capability to solve medical emergencies may guide the design of future onboard healthcare systems.
	
	We present ten basic principles and concept design of a software suite to bring onboard decision support to help the crew dealing with medical emergencies taking into consideration physiological disturbances in space and spaceflight restrictions.
	1) give real-time support for emergency medical decision making, 2) give patient-specific advice for executive problem-solving, 3) take into account available information from life support and monitoring of crewmembers, 4) be fully autonomous from remote facilities, 5) continuously adapt predictions to physiological disturbance and changing conditions, 6) optimize emergency medical decision making in terms of mission fundamental priorities, 7) take into account medical supplies and equipment on board, 8) apply health standards for the level of care V, 9) implement ethics responsibilities for spaceflights, and 10) apply ethical standards for artificial intelligence.
	
	We propose an autonomous clinical decision support system (CDSS) to provide real-time advice for emergency medical interventions on board of space exploration missions.
	The software suite is conceptually designed as four interconnected modules. The main of them is responsible for giving direct advice to the crew by predicting the medical emergency characters (life-threatening, delayability, ethical dilemma, duration of therapy, and compatible diagnoses), classifying the required tertiary medical intervention and optimizing the medical action plan. This module's output is continuously evaluated and re-trained with changing physiological data from the crew by an adaptive learning module, ensuring fairness, interpretability, and traceability of decision-making during the computational system's full operational time. Finally, the suite interacts with the onboard health information systems using a mapping layer to semantic interoperability standards.
	
	The installation of clinical decision support systems on board of future missions to Mars may facilitate a comprehensive preventive emergency medical strategy. Moreover, the advance in technology may represent a stepstone of the future quantitative medicine on Earth and the expansion of humans throughout the Solar System.
	
\end{abstract}

	%\scalebox{1}{\includegraphics*[width=\columnwidth,natwidth=3421,natheight=3672,angle=0]{figures/MEDEA_graphicalabstract}}
	

%	\item Emergency healthcare based on stabilize and transport is not applicable to Mars missions
%	\item Ten principles to design onboard clinical decision support systems are discussed
%	\item A conceptual design is presented based on prediction modeling, adaptive learning, ethics proactivity and semantic interoperability

%\begin{keyword}
%	human spaceflights \sep space exploration \sep Mars \sep clinical decision support systems \sep deep/machine learning \sep emergency medicine \sep space medicine \sep adaptive learning \sep semantic interoperability \sep ethics
%\end{keyword}

%\end{frontmatter}

\section{Introduction}

%- Tres expedicienes robóticas están viajando actualmente hacia Marte aprovechando la ventana de cercanía entre los planetas. La siguiente ventana serán en 2035 y las agencias espaciales y compañias privadas preparan sus próximas misiones, que posiblemente pondrán a la humanidad en otra planeta por primera vez en la historia.
%- Carrera espacial centrada en Marte. Varias agencias espaciales han puesto objetivo en planeta rojo como paradigma de mision de exploracion espacial. Enumerar objetivos y programas: asteroides y marte. Si esta ventana se ha aprovechado para lanzamiento de exploradores rover: perseperance, arabe, ... la proxima ventana de viaje corto 2035 posria ser el momento de ls primera mision tripulada a Marte.

In early 2021 three spacecraft, Mars 2020, Tianwen-1 and Hope, have arrived at Mars with the mission of studying its atmosphere and surface with the potential result of detecting clear signs of life~\cite{witze_countdown_2020}.
The three spaceflights are exploration missions based on robotics, but it is not the orbital window for human exploration of Mars yet.
NASA, other governmental space agencies and private companies prepare the beginning of the planetary-type missions with humans for the 2030s transfer windows~\cite{wooster_mission_2007}, intending to put the first human on the Mars surface after achieving the moon again through its orbital Gateway station~\cite{smith_artemis_2020}.
This new achievement may open a new era of deep space journeys and planetary design reference missions, including long-term stays on Mars, Lagrange points and exploitation of near-earth asteroids~\cite{united_statescongresssenatecommittee_on_commerce_science_and_transportation_future_2011}, opening unseen opportunities to humanity.

% - el diseño viajes de exploración espacial con humanos y exploracion a marte deben cambiar complementamente desde el punto de vista de salud y performance de la tripulación y posibles pasajeros.
\medskip

Although exciting, this challenge with not be easy to achieve for humankind given that the absence of gravity and radio-protective geomagnetic fields from Earth, along with distance, isolation and hostile environments experimented during long-duration space travels, may increase systematic effects on the physiology, biology and behavior compared to those measured during short-term experiments on the International Space Station (ISS) in a low Earth orbit~\cite{grimm_guest_2019}.
Then, a review by NASA~\cite{romero_nasa_2020} of thirty human health and performance risks for space exploration pointed out to increase crew autonomy to manage inflight medical conditions.

%sobre riesgos de salud humana el impacto para la mission
Many of the consequences of human risks in long spaceflights are not fully understood yet, and technologies for controlling them are still to be invented. 
NASA established the Human Resource Program at Johnson Space Center in 2005 for investigating the highest risks to astronaut health and performance by quantifying the likelihood of occurrence, the severity of consequences, and the extent that risk can be controlled or mitigated both inflight and post-flight for each type of mission.
Indeed, its path to risk reduction for a planetary mission estimates that {\em inflight medical conditions} with potential high impact~\cite{noauthor_nasas_2015} will only be partially controlled in time for a Mars mission in the 2030s.

\medskip

The accepted approach for the crew's healthcare on a space mission follows the occupational medicine prevention strategy~\cite{hamilton_autonomous_2008}.
It focuses on reducing the likelihood and severity of medical events by primary, secondary and tertiary interventions.
With primary interventions, the likelihood of risk factors is reduced by careful selection of crewmembers.
For example, priority is currently given to astronauts with low coronary artery calcium and low Framingham risk scores over those with higher risk levels.
Besides, secondary interventions are also applied as countermeasures for the effects of environmental factors in space.
In low Earth orbit missions, countermeasures to the effect of microgravity and isolation are carried out by routines on treadmills, resistive devices and cycles.
Finally, tertiary interventions are activated to treat illness or injury in emergency situations requiring specific types of care.
Given the difficulty to provide full healthcare support to space, current low Earth orbit and lunar missions have followed the paradigm {\em stabilize and
transport} the injured crewmember to a terrestrial definitive medical care facility that is not applicable to exploration class missions.

\medskip

%restricciones de diseño
Deploying a medical strategy for controlling inflight medical emergencies in exploration class missions will need to deal with the limitations imposed by deep space hazards.
Three are the major limitations affecting the healthcare of mission crewmembers: physiological disturbance, communication latency and mission length:

\begin{itemize}
\item Physiological disturbance includes radiation-induced changes, altered nutritional status, neurovestibular deconditioning, cardiovascular deconditioning, bone and muscle loss, renal-stone formation, plasma-volume shifts, spaceflight-associated neuro-ocular syndrome, and altered immunity due to microgravity, radiation and isolation, among others~\cite{stepanek_space_2019}.
Multiple biomedical experiments have been performed on board the ISS~\footnote{https://nebula.esa.int}.
For example, the NASA Twins Study identified multiple onboard spaceflight-specific changes, including decreased body mass, telomere elongation, genome instability, carotid artery distension and increased intima-media thickness, altered ocular structure, transcriptional and metabolic changes, DNA methylation changes in immune and oxidative stress-related pathways, and gastrointestinal microbiota alterations~\cite{garrett-bakelman_nasa_2019}.
Nevertheless, there is no evidence of the effects of long-distance travels on humans yet.
Hence, healthcare strategies, including clinical decision making, will need to be adaptive to unseen changes, such as new clinical conditions, symptoms, and complications.
\item Communication latency between the crew and mission control will increase with distance to Earth.
From three to six seconds delay for a round trip communication from ISS to Earth, any deep space journey to near-earth asteroids and planetary missions will take several minutes for full bidirectional communication.
This precludes asking immediate inquiries to a telemedicine service similar to what ISS uses to consult a mission surgeon at mission control.
Hence, autonomous real-time decision-making based on available onboard health information will need to evaluate when a medical emergency is not delayable and how to proceed following clinical, ethical, and legal rules applicable to the space mission.

\item The lengths of the space missions will be extended from a current maximum of months to several years. 
The primary condition that mission length implies is the non-return to permanent health facilities in case of requiring advanced healthcare. Therefore, current mission designs by NASA approach medical care by high levels of self-sufficiency~\cite{drake_human_2009}.
Moreover, due to the mission length, we may experience cumulative effects in the physiological disturbance of human beings, so the increasing probability of health failure and clinical severity with time is expected.

\end{itemize}

%cambio de paradigma
As Hamilton {\em et al.} stated on~\cite{hamilton_autonomous_2008}, the current tertiary intervention can no longer draw on a close definitive medical care facility at an effective time, so the medical design should evolve into an {\em autonomous treat to final resolution capability}, which represents a significant challenge to space medicine and mission designers.
In this paper, we tackle the challenge of managing medical emergencies in Mars-type missions by proposing a {\em real-time} {\em clinical decision support system} based on {\em adaptive learning}.
For that, our study establishes ten basic pillars and proposes a concept design consisting of four functionalities: {\em autonomous real-time clinical decision making}, {\em space adaptive learning}, {\em semantic interoperability} and {\em ethical \& legal functional support}.

We are not aware of any scientific paper or experiment report about information technology specialized in emergency medicine in space, being this study the first that propose basic principles and a concept design of clinical decision support systems to tackle autonomous tertiary medical interventions.
This concept design answers the system-based strategic vision conceived by Williams, Hamilton, Doarn and others~\cite{williams_strategic_2000, hamilton_autonomous_2008, doarn_medical_2010} utilizing the last decade advances in Biomedical Data Science and Artificial Intelligence.

\section{State of the art}

%%% estudios estrategicos de programas espaciales
The routine and emergency medical operations in the human flight missions are currently managed by the flight surgeons from the mission control center.
Crewmembers have an onboard checklist~\cite{mission_operations_directorate_international_2001} that describes the routine and emergency medical operations procedures and hardware associated with crew health.
Nevertheless, these procedures based on telemedicine and prompt evaluation for tertiary medical interventions should change for exploration and planetary missions.
The design of Mars missions includes re-thinking the life support and medical management for remote human risks management.
In this way, American and European space programs have detected unresolved technologies to manage medical interventions on board of spaceflights in long exploration missions such as Mars-type missions.

The Human Research Program by NASA detected the need of having the {\em capability to provide computed medical decision support during exploration missions}\footnote{https://humanresearchroadmap.nasa.gov/gaps/gap.aspx?i=642} as part of the detected gap {\em to identify new capabilities that maximize benefit and reduce costs on the human system/mission/vehicle resources}\footnote{https://humanresearchroadmap.nasa.gov/gaps/gap.aspx?i=716}.
By solving this gap, NASA is willing to develop continued monitoring of biomedical signals and images, improve medical capability technology for unique spaceflight needs, provide medical care in a progressively Earth independent fashion and demonstrate the integration of the new procedures and technologies with the onboard processes.

%EU - ESA
%THESEUS
Besides, the European strategy towards the human exploration of space developed during the THESEUS FP7 project in 2012~\cite{noauthor_towards_2012} detected {\em insufficient onboard expert systems/decision support systems for medical diagnostics} as the specific key issue for the space medicine expert group. %%%% sobre telemedicina
%assist
In 2013, the ASSIST report by ESA/ESTEC encountered a major limitation in the data gathering for remote monitoring by telemedicine. Specifically, they reported the bottleneck of acquiring and transmitting too much data not always immediately and easily available and the difficulty to interpret them, so 75\% of validators considered it useful to incorporate additional functionalities such as the automatic identification of indicators with high sensitivity and the automatic identification of values for the break-even.
%AMIGO
Then, CSEM, Airbus and MEDES evaluated in 2016 the benefit of data mining for an autonomous medical monitoring/diagnostic system (AMIGO) in long-term spaceflights and non-space related applications.
Their studies estimated higher medical risk due to expected human extravehicular activities during asteroid missions for sample extraction and planetary surface exploration for the Mars-type missions.
They performed two proofs of concept on subjects with {\em de novo} cardiac arrhythmia and sleep apnea, respectively. For their computational experiments, they applied linear classifiers, quadratic classifiers, Gaussian mixture models, hidden Markov models, artificial neural networks, k-nearest neighbors, decision trees, Bayes networks, and random forest to biomedical signals from the Physionet open repository\footnote{https://physionet.org}.
%, characterized by high occurrence (3 and 0.15 person/year during transfer and planetary/asteroid surface activities respectively), especially when the subject is under stress environment and microgravity with a relative dehydration/hemoconcentration as observed on astronaut after some days under microgravity 
%o Sleep apnea, characterized by high occurrence (10% of an astronaut during 1 to 3 days at gravity changes during transfer and planetary/asteroid surface activities, respectively 
% Databases: De Novo cardiac arrhythmia (MIT-BIH Arrhythmia, MIT-BIH Noise Stress, CU Ventricular Tachyarrhythmia, Long-Term AF Databases) 94 subjects, 30'..24h. multi-lead ECG recordings (from 2 to 12-lead ECG) with useful medical- condition-dependent features and classification. Sleep apnea syndrome (MIT-BIH Polysomnographic Database) Monitoring duration: from 97 minutes up to 6 1/2 hours Population: 16 subjects (18 recordings), ECG, invasive blood pressure signal, EEG, respiratory signals (EOG, EMG, stroke volume and oxygen saturation are optional)

\medskip
%%% CDSS
%McGregor, Artemis, IBMP, HRV, health state gravities, Cosmocard, ISS
McGregor in 2013~\cite{mcgregor_platform_2013} proposed a real-time platform call Artemis for online health analytics during spaceflight by monitoring the astronauts' physiological signals as well as sending the signals to mission control for medical support at each stage when communication was available.
Prysyazhnyuk {\em et al.} in 2017~\cite{prysyazhnyuk_big_2017} tested by analogs if Artemis was able to support the implementation of the classification model of health states gravities in four functional states (physiological normal, prenosological, premorbid and pathological) based on the discriminative Heart Rate Variability defined by the Institute of Biomedical Problems of the Russian Academy of Sciences.
Moreover, in 2017 McGregor and the Institute of Biomedical Problems~\cite{mcgregor_method_2017} simulated the integration of the Artemis platform and the Cosmocard device\footnote{https://www.energia.ru/en/iss/researches/human/12.html} acquiring electrocardiograms from the Russian cosmonauts on ISS. Their results have shown limitations for a real-time performance due to deferred transmission from the Cosmocard device.

\medskip

%%%%biomedical signals
In addition to medical diagnostics, biomedical signals have been acquired on board of the ISS to study the effects of microgravity and isolation in human physiology.
EveryWear by ESA monitored electrocardiography, tonometry, and temperature of astronaut Thomas Pesquet from Nov 2016 to May 2017 by wearable sensors connected to an iPad.
The Neurolab Spacelab mission in 1996 studied the effects of weightlessness on the brain and nervous system~\cite{bondar_neurolab_2005}.
Petit {\em et al.} in 2019~\cite{petit_local_2019} studied the relationship of sleep pressure electroencephalography markers during wakefulness in astronauts throughout a 6-month space mission.
Gemignani {\em et al.} in~\cite{gemignani_pattern_2015} performed what we may consider the first adaptive data-driven decision support system for space medicine.
They defined dynamical thresholds on high-density electroencephalography to detect sleep spindles in variable sleep depths and subjects.
Besides, the Airway Monitoring ISS investigation studied the inflammation and reduced pressure on pulmonary nitric oxide turn-over due to microgravity and 
other ESA experiments such as DNAmAGE, ICELAND and IMMUNO have also studied spaceflights' effect at the genomic, microbiome and immunological levels.
%DNAmAGE - Effects of prolonged spaceflight on DNA methylation age 2020 • Space Stations • ISS Increment 63 • ID: 9838 (performed the 1st time) A. Søraas, S. Horvath, K. Raj
%4.9 ICELAND - Immune and microbiome Changes in Environments with Limited ANtigen Diversity 2019 • Ground Facilities • CONCORDIA winter-over 2019 • ID: 9743 (follow-up) P. Enck, J. Doré, C. Lambert, J. Penders
%4.9 IMMUNO 2 - Consequences of Stress Challenges on stress response systems and Immunity in Space: a multidisciplinary approach 2019 • Space Stations • ISS Increments 61-62 • ID: 9516 (follow-up) A. Chouker, S. Baatout, M. Morrels, R. Quintens, B. Crucian, C. Sams, D. De Quervain, M. Feuerecker, I. Kaufmann, G. Schelling, D. Hauer, M. Hörl, S. Matzel, J.I. Pagel, N. Montano, E. Tobaldini, B. Morukov, I. Nichiporuk, M. Rykova, V. Gushin, S. Praun, B. Roozendaal, M. Thiel, P. Campolongo
%4.10 Physical Performance and Biological Age 2019 • Ground Facilities • AG-BR-ESA long-term bed rest DLR-ESA-NASA - :envihab 2019 • ID: 9769 (performed the 1st time) M. Gruber, A. Kramer, A. Bürkle, M. Moreno-Villanueva, C. Franceschi, E. Jansen, T. Grune

\medskip

%%%% medical image
Up to now, ultrasonography is the main onboard diagnostic imaging technique. ESA's Downstream Gateway project\footnote{http://youbenefit.spaceflight.esa.int/ultrasonography-without-borders/} and the ECHO experiment tried to solve the operator-dependence by remote-controlled ultrasounds operations. Although this solution is feasible for low Earth orbit missions, long latency in exploration and planetary missions will require autonomous onboard operation. 
%raspberry pi for diagnosis: visual impairment due to intracranial pressure
%Aravindhan {\em et al.} proposed in~\cite{aravindhan_medical_2020} a Raspberry Pi solution for online health diagnosis to operate during space tourism and future Mars colonization by illustrating its potential application to fundoscopies when suspicion of visual impairment due to intracranial pressure.
%Moreover, these projects are centered on solving specific space medical problems instead of managing medical emergency situations.
% colocar donde vaya
%no es MRI on-board, solo estudios simulados
%4.8 BRAIN-DTI - \cite{van_ombergen_spaceflight-induced_2017} Spaceflight induced neuroplasticity studied with advanced magnetic resonance imaging methods

\medskip

%%%risk management
The Human Research Program has identified 32 physiological, medical, and behavioral risks associated with long-duration spaceflights~\cite{stewart_emergency_2007}.
Linked to that, Davis {\em et al.} presented in~\cite{davis_human_2008} a risk management system based on the acceptable levels of risk for each mission type to guide research efforts and mission planning through the probability of adverse medical events, the uncertainty of outcomes, impacts, costs and benefits of mitigation actions along with related current and future work.
More recently, Mindock {\em et al.} in~\cite{mindock_integrating_2017} defined a connected map of contributing factors and the medical risks. In contrast, Romero and Francisco in~\cite{romero_nasa_2020} identified one hundred probable health problems that may affect mission success and classified the medical risks in five hazards of spaceflight: altered gravity, radiation, distance, isolation and hostile environment.
Taking into account that medical care will be limited by mass, volume, and power constraints and that life support will represent the 40\% of wet mass in exploration-class spaceflight~\cite{gemignani_beyond_2015}, one of the firsts uses of the risk assessment presented above was to generate a list of onboard medical resources.
With that objective in mind, Antonsen {\em et al.}~\cite{antonsen_prototype_2018} designed a tradespace analysis tool to score resources, tools, and skillsets required for exploration missions.

\medskip

%As we have seen, current research is focused on identifying adaptation effects on spaceflights.
Whereas biomedical signal monitoring is almost routinary in astronauts, current solutions are solved by telemedicine. As a result, few advances have been made on autonomous decision making.
Besides, medical risks, diseases, factors, mitigations, and consequences identified in healthcare management studies constitute the key knowledge to plan research directions for designing human spaceflight missions to Mars and asteroids.
Hence, our proposal of the clinical decision support system focuses on the specific requirements that crewmembers will have to deploy autonomous tertiary medical interventions
for managing medical emergencies.

\section{Basic principles of onboard decision support systems for managing medical emergencies\label{sec:principles}}

Primary and secondary medical interventions for Mars-type missions can be operated in advance and from mission control, respectively, as they are currently deployed for ISS missions.
Nevertheless, long-distance exploration and Mars missions require autonomy to decide how to react to a medical situation~\cite{drake_human_2009}.
To test the need for decision support systems for medical emergencies on board of exploration missions, we inspected the one hundred health conditions listed by Romero and Francisco identified in~\cite{romero_nasa_2020} as potential problems that may compromise the mission success.
For that, we asked the opinion of three independent medical doctors with experience in medical emergencies.
Specifically, each health condition was classified as potential life-threatening situation (Yes/No), if the patient would need medical attention before 15 minutes and if the condition may suddenly appear in healthy people although they are monitored.
As described in Table~\ref{tab:romerosisisisi}, medical doctors identified from 23 to 48 life-threatening situations in the Romero and Francisco list. From the one hundred conditions, from 31 to 42 were considered non-delayable health problems, and from 40 to 86 would suddenly appear in healthy people although they are monitored.

Moreover, medical doctors classified from 14 to 32 health problems as sudden life-threatening situations that may require medical attention before fifteen minutes, being the eight medical conditions included in Table~\ref{tab:romerosisisi2} those where the three doctors achieved consensus.
We have also included in the table the most relevant type of medical decisions that each medical condition may require: differential diagnosis, cause, further diagnoses, prognosis and treatment planning.

\begin{table}
\caption{Number of health problems on board of exploration missions identified by Romero and Franscisco~\cite{romero_nasa_2020} that may represent sudden life-threatening situations requiring medical attention before 15 minutes.\label{tab:romerosisisisi}}
\footnotesize
%\begin{tabular}{p{4cm}|p{2cm}p{2cm}p{2cm}p{2cm}|}
\begin{tabular}{p{6cm}c}
	\hline
	Risk & Number of medical conditions [range]\\
	\hline
	Life-threatening situation & [23, 48]\\
	Delayability $< 15'$	& [31, 42]\\
	Sudden although monitored	& [40, 86]\\
	%Life-threatening and demorability $< 15'$	& [14, 32]\\
	Sudden life-threatening with delayability $< 15'$ & [14, 32]\\
	\hline
	Consensus: sudden life-threating situations with delayability $< 15'$ & 8\\
	\hline
\end{tabular}
\end{table}

\begin{table}
\caption{Health problems on board of exploration missions identified by Romero and Franscisco~\cite{romero_nasa_2020} that may represent sudden life-threatening situations requiring medical attention before 15 minutes. Columns indicate what types of decision may require each medical condition: {\em differential diagnosis}, {\em cause}, {\em further diagnoses/prognosis} and {\em treatment planning}. \label{tab:romerosisisi2}}
\footnotesize
%\begin{tabular}{p{4cm}|p{2cm}p{2cm}p{2cm}p{2cm}|}
%\begin{tabular}{p{5cm}|p{2.5cm}p{2.5cm}p{2.5cm}p{2.5cm}}
\begin{tabular}{ccccc}
	\hline
	Health problem & Differential & Cause & Further diagnoses &Treatment\\
	& diagnosis & & and/or Prognosis & planning\\
	\hline
	Angina/Myocardial infarction	& Yes	& Yes	& Yes	& Yes\\
	Anaphylaxis & No & Yes & Yes & Yes\\
	Chest injury & No & No & Yes &Yes\\
	Electric shock injury & No & No	& Yes &Yes\\
	Neurogenic shock & Yes &Yes &Yes & Yes\\
	Seizures	& No & Yes & Yes	& Yes\\
	Sudden cardiac arrest & Yes & Yes & Yes & Yes\\
	Traumatic hypovolemic shock & Yes &	No	&  Yes & Yes\\
	\hline
\end{tabular}
\end{table}

\medskip

Intending to give the highest standards of healthcare decision making during an exploration mission to Mars, we study how to bring onboard computational support to manage emergency care situations.
As we have seen, for those medical situations that represent sudden (or previously unseen) life-threats requiring an urgent medical response, the exploration mission would need to act autonomously from mission control.
Then, onboard resources for predicting, deciding and planning healthcare may be critical elements to save lives.
With that in mind and given the limitations imposed by deep space hazards, we propose and argue the next ten basic principles for designing onboard clinical decision support systems for deep exploration and Mars missions:

\begin{enumerate}
\item Give real-time support for emergency medical decision making

Giving support during medical emergencies includes evaluating the character of the medical emergency, predicting compatible diagnoses, classifying the required tertiary care intervention and planning healthcare actions. Evaluating a medical emergency character includes classifying the event as a life-threatening situation or not, along with delayability in assisting the patient.
Besides, deciding the type of tertiary care intervention required for managing the patient may guide the action plan subject to the available supplies and equipment on board.
% (basic life support, first-aid, ambulatory care, advance cardiac life support, advanced trauma life support, basic surgical care and palliative care)

\item Give patient-specific advice for executive problem-solving

More than policies, standards, procedures, manuals of health systems and medical checklists, the concept design of the software suite may include direct support for decision making for emergency healthcare of individuals.
Then, two main elements guide the interaction of the system with the users.
First, the system may react proactively to the health situation according to the mission's healthcare procedure. 
Second, the system may be fully informed of the mission's healthcare status, keeping records of the crewmembers' quality of life, environmental conditions, and remaining supplies.

\item Take into account available information from life support and monitoring of crewmembers

Current technology allows combining data from different nature for prediction and decision making.
Then, time series of physiological signals, nutrition and digestion tables, adverse events, air and water quality, psychological status can add value to the health situation by structured (quantitative description of the patient status) and non-structured (voice and free text notes). Moreover, electronic health records with crewmembers' history may complement the relevant information to give patient-specific advice.

\item Be full autonomous from remote facilities

Long-distance space exploration missions from Earth imply no return to a terrestrial medical care facility and transmission latency of several minutes with mission control. 
This requires autonomy of the mission to do healthcare interventions and capability to detect and decide management of life-threatening situations that are not delayable.
A real-time clinical decision support system may give processed knowledge related to patients' conditions during onboard emergency situations.
In extreme situations, for missions designed with one doctor on board, the system may be ready to assist crewmembers in the critical case Chief Medical Officer is unavailable or indisposed. 

\item Continuously adapt predictions to physiological disturbance and changing conditions

Decision making in medical emergencies must be aware of physiological disturbances due to the lack of gravity, radiation-induced changes, altered nutritional status, neurovestibular deconditioning and cardiovascular deconditioning, among other effects.
This continuous change in the presentation and prevalence of diseases may generate a dataset shift in the crewmembers' data.
Therefore, a continuous adaptation of prediction and decision-making must correctly interpret the observations during medical emergencies.

\item Optimize emergency medical decision making in terms of mission fundamental priorities

Hamilton {et al.}  in~\cite{hamilton_autonomous_2008} compiled the mission fundamental priorities from NASA medical mission standards, so they ordered vehicle survival, health \& safety of the crew, mission success, and payload success in descending order of importance.
We can then expect any system supporting emergency medical decision-making to comply with these (or future refined) mission fundamental priorities.

\item Take into account  onboard medical supplies and equipment

Given the high cost of medical payload, crewmembers will have access to a limited amount of supplies, equipment and resources during the mission. Hence, optimization of prescriptions will be needed on every day-by-day medical decision making.

\item Apply health standards for the level of care V

Any system supporting emergency medical decision making would have to comply with the space flight human-system standards applicable to the mission. In its Space Flight Human System Standard~\cite{williams_nasa_2015}, NASA defined the levels of care that one may provide depending on the mission type.
Levels from zero to five are defined depending on several factors: distance, duration, health risks, and available technologies to assign the level to a mission.
Exploration missions are assigned to Level of Care V, given that return to Earth is not a viable option for more serious illness and injuries, representing a potential overall impact on the mission.
Hence, exploration missions, and Mars missions, in particular, may entitle the most complete set of medical capabilities, including basic and advanced life support, first aid, clinical diagnosis, imaging, ambulatory care, telemedicine, sustainable advanced cardiac life support, advanced trauma life support, basic surgical care and palliative care~\cite{hamilton_autonomous_2008}.

\item Implement ethics responsibilities for spaceflights

Implementing a system for medical decision making may recognize ethics responsibilities identified relevant for human health on long-duration and exploration spaceflights, such as some of those described by Kahn {\em et al.} in~\cite{kahn_recommendations_2014}: a) fully inform crewmembers about risks to allow informed decision making, b) continuously improve decision making by adopting knowledge gained from data gained during missions, c) provide preventive lifetime healthcare decision making, d) protect privacy and confidentiality of crew health data.
%Emphasis on the confidentiality of astronaut clinical data has resulted in missed opportunities to understand human physiological adaptations to space~\cite{ball_safe_nodate}.

%Another relevant issue is the application of ethical criteria when giving real-time decision support for solving medical events.
%To this, fundamental mission priorities and available resources for Level of Care V may guide the medical action plan's optimization.

\item Apply ethical standards for Artificial Intelligence

Ethics responsibilities for spaceflights must be complemented with those described for artificial intelligence frameworks.
Given that we are proposing solutions based on data-driven decision support systems, independently of the medical certificates requested to the crew medical officers~\cite{stepanek_space_2019}, our proposal may also adopt solutions to ensure a) fairness to avoid reproducing any pattern of discrimination due to prejudices or bias~\cite{oliver_governance_2019}, b) interpretability to enable a correct explanation of medical predictions and decisions by human experts~\cite{reyes_interpretability_2020}, and c) traceability to achieve a comprehensive examination of responsibilities of any medical decision making at any time, ensuring the currency of the knowledge base and that it is safe to use~\cite{shortliffe_clinical_2018}.

%  First, the massive use of biomedical data of the crewmembers in the context of a unique environment of deep space requires the correct management of regulation for data protection and privacy.
%Laws adopted to consider the digital era, such as the General Data Protection Regulation (GDPR)~\cite{noauthor_regulation_2016} implemented by the European Union in 2018, give data control to the individual.
%Our premise is that the new regulations for data protection may be a mechanism to avoid the loss of valuable biomedical information from deep space environments.

%\item Be in the position to pass any audit from regulation authorities about decision making and data re-use 

\end{enumerate}

\section{Concept design of MEDEA\label{sec:design}}

\begin{figure}
\centering
%\scalebox{0.9}{\includegraphics*[width=\columnwidth,natwidth=4980,natheight=4788,angle=0]{figures/MEDEA_concept}}
\scalebox{0.08}{\includegraphics*{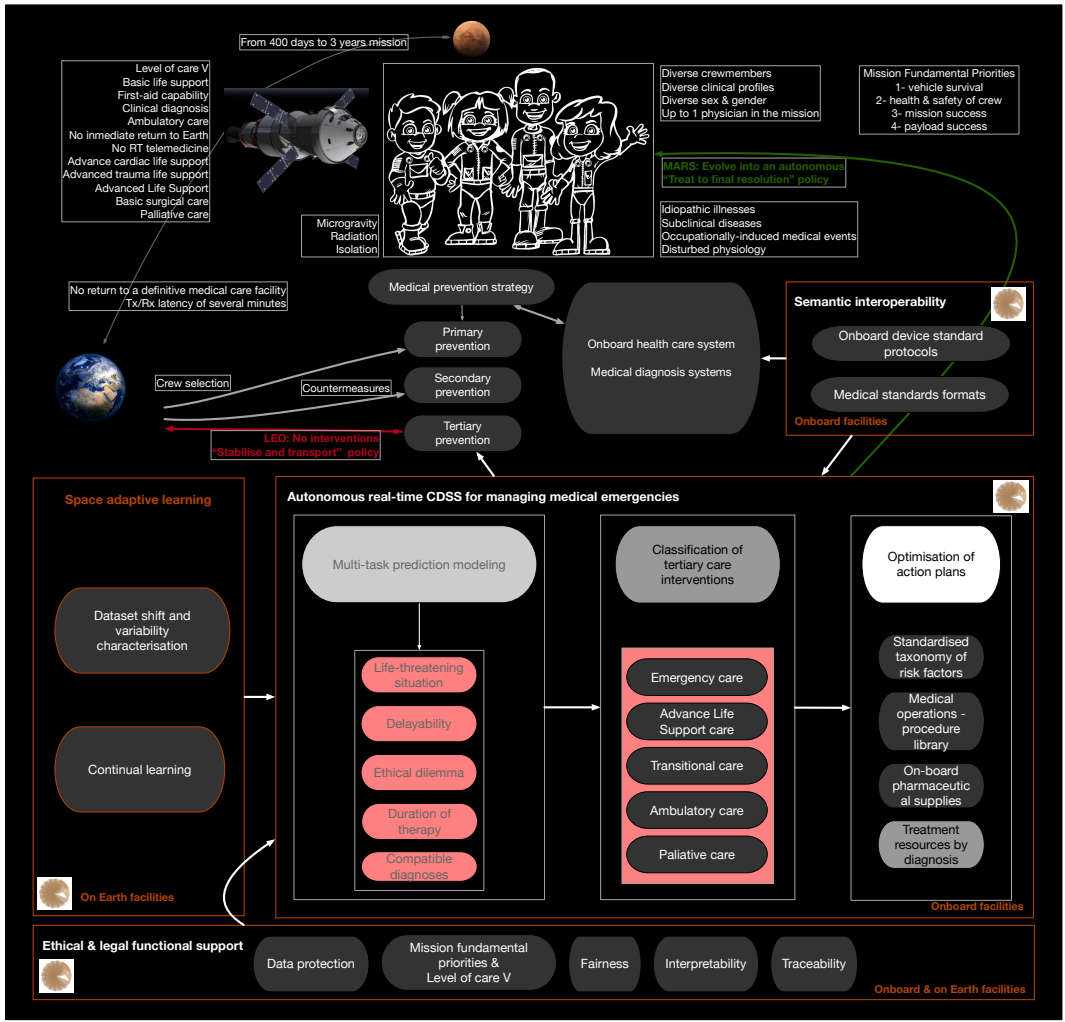}}
\caption[Concept design of MEDEA.]{Concept design of MEDEA composed of four main subsystems for {\em autonomous decision making}, {\em space adaptive deep learning}, {\em semantic interoperability} and {\em ethical \& legal functional support}.\label{fig:medea_concept}}
\end{figure}

We propose MEDEA as a comprehensive computational suite to deploy personalized clinical decision support for medical emergencies during exploration and planetary missions to illustrate the specific modules needed to achieve the ten basic principles described in the section above.
Such software should operate continuously to react to emergencies and unseen medical situations to advice on clinical decision making by quantitative predictions adapted to individual profiles.
Our design is composed of four software modules physically distributed on board and on Earth facilities to provide four main functionalities of the system: {\em autonomous decision making} (onboard), {\em space adaptive learning} (on Earth), {\em semantic interoperability} (onboard) and {\em ethical \& legal functional support} (onboard and on Earth). Figure~\ref{fig:medea_concept} shows the details of how MEDEA supports tertiary interventions during medical emergencies and how the modules are interconnected among them and with the  onboard information systems.

The module {\em Autonomous real-time CDSS for medical emergencies} is an artificial intelligence computational engine that directly interacts with the crewmembers to support medical decision-making under the paradigm {\em treat to final resolution}.
To do that, it receives the health status of the crewmembers from the onboard health care system and the medical diagnosis systems through the {\em semantic interoperability} module.
The artificial intelligence models are continuously adapted to the changing space conditions by the {\em space adaptive learning} module. 
Both CDSS and deep learning performances are continuously verified by the {\em Ethical and legal functional support} module.
%In the next four sections, we provide the technological solution of each module of MEDEA.

\section{Autonomous real-time CDSS for tertiary medical care}

Romero {\em et al.} in~\cite{romero_nasa_2020} compiled the most common hundred medical conditions in space derived from the ISS Medical Checklist, scientific research and occupational health statistics. This list's stratification serves the Human Research Program of NASA for planning mitigation actions for the thirty human system risks.
Medical conditions in spaceflights may be occupationally-induced conditions or idiopathic illnesses.
Crew activities confined in a spacecraft extravehicular activities and surface explorations may increase the probability of injuries and trauma that may derive from emergency medical situations.
Although presentations and frequencies of medical conditions during prolonged stays may change for Mars missions, the closest reference given by medical reports from ISS~\cite{crucian_incidence_2016} revealed that 46\% of crewmembers expressed an event deemed "notable", being skin rashes and hypersensitivities (40\%, 1.12/flight year) along with upper respiratory symptoms (0.97/flight year), the most reported events.
Moreover, artificial life support added to space-specific conditions and isolation increase the onset of conditions such as space adaptation syndrome, headaches, gastrointestinal distresses, degradation of the immune system, infectious processes, sleeplessness and depression, among others.

On top of that, idiopathic illnesses during 3-4 years of a Mars-type mission are willing to appear more than in a low Earth orbit mission due to: prolonged stays of the same crewmembers, the variability of tasks during exploration missions~\cite{ahlf_mars_2000}, a complete absence of gravity, exposure to radiation and increase in the number of astronauts from private and public space programs and higher variability of medical profiles~\cite{stepanek_space_2019}.

The current design of exploration-type and Mars-type missions plans to book one seat of the spacecraft for a physician acting as Chief Medical Officer in every spaceflight crew~\cite{kuypers_emergency_2013,hamilton_autonomous_2008}.
While this may improve mission safety, the long latency to communicate from mission control means isolation in medical decision-making to resolve emergencies.
A real-time clinical decision support system may give processed knowledge related to patients' conditions to the Chief Medical Officer during onboard emergency situations.
Moreover, the system's result may assist crewmembers in the critical case Chief Medical Officer is unavailable. 

Furthermore, the design of MEDEA should also follow the currently accepted caveats for clinical decision support systems by Shortliffe and Sepulveda~\cite{shortliffe_clinical_2018}: a) black boxes are unacceptable, b) complexity and lack of usability thwart use, c) delivery of knowledge and information must be respectful, and d) scientific foundation must be strong.

As a result, we consider this module may perform three sequential functionalities to give complete support for dealing with emergency medical situations: multitask prediction modeling, classification of tertiary medical interventions and optimization of action plans. Each functionality and its possible design based on current technology are described below.

\subsection{Multitask prediction modeling~\label{sec:multitask-prediction-model}}

When a medical emergency arises, a prompt prediction should be carried out to decide \#1 if it is a life-threatening situation, \#2 its delayability, and \#3 if it represents an ethical dilemma for the mission fundamental priorities, along with the estimation of the \#4 duration of therapy and \#5 the list of diagnoses compatible with the situation.
A positive prediction of a life-threatening situation may activate onboard emergency protocols that, depending on the estimated delayability, may wait for advice from mission control to act or not.
Moreover, when a situation is considered an ethical dilemma, specific constraints may be contemplated from the beginning of the action plan, taking into account the treatment's potential duration.
Besides, the ordered list of compatible diagnoses may guide the specific set of medical operations to restore the crew's health and performance.
Current approaches such as the deep neural network with hard parameter sharing~\cite{ruder_overview_2017, caruana_multitask_1997} for emergency medicine proposed by Ferri {et al.} in~\cite{ferri_deep_2020} may be optimal to exploit the dependencies among the five prediction tasks.

From the five prediction tasks, \#1 life-threatening situation and \#3 ethical dilemma are classification tasks (i.e. yes/no), \#2 delayability and \#4 duration are regression tasks (i.e. positive numbers in a limited range), and \#5 is a recommendation task (i.e. ordered list of diagnoses).
For Task \#5 compatible diagnoses, the list of potential diagnoses may be large. 
A total of 71932 codes of diagnoses are included in version 10 of the International Statistical Classification of Diseases and Related Health Problems (ICD) by the World Health organization.
From them, designers of Mars missions {\em should plan to treat} those diagnoses affected by 135 conditions detected by Nusbaum {\em et al.} in 2019~\cite{nusbaum_development_2019}.
%Additional 22 conditions were {\em plan to treat with conditions} and 34 of them {\em should not plan to treat}.
Hence, one feasible configuration for Task \#5 is a recommendation system~\cite{su_survey_2009} that, given the medical situation, returns an ordered list of compatible diagnoses from the overall set of potential diagnoses.

   %    \medskip
%Current approaches such as the deep neural network with hard parameter sharing~\cite{ruder_overview_2017, caruana_multitask_1997} for emergency medicine proposed by Ferri {et al.} in~\cite{ferri_deep_2020} may be optimal to exploit the dependencies among the five prediction tasks.
%Following the Ferri's proposal, a possible architecture for our multitask prediction modeling may ensemble four task-specific subnetworks with specific parameters for tasks \#1 to \#4 and a task-shared subnetwork sharing a set of parameters for all prediction tasks.
%The task-shared subnetwork may be a dense block composed of fully connected layers, a batch normalization layer, a leaky ReLU~\cite{maas_rectifier_2013} activation function and a dropout layer.
%The task-shared subnetwork output is the input of the five task-specific subnetworks, all of them configured as dense blocks but with different output layers. Soft-max functions are used for classification tasks \#1 and \#3, whereas linear functions are used for regression tasks \#2 and \#4.
%Additionally, following the multi-view deep learning approach by~\cite{elkahky_multi-view_2015}, we can obtain a latent space for task \#5 recommendation system from the fully connected layers of the task-shared subnetwork.
%Then, an ordered list of diagnoses is obtained by the nearest neighbors' votes (training cases) of the test case in the latent space.
%\medskip

As described above, the module's input is expected to be paramount of biomedical data from crewmembers and situation awareness.
Structured information from electronic health records may be combined with free-text reports and speech notes of the emergency scenario.
Previous conditions and patient evolution may be extracted from longitudinal signals acquired from wearable monitoring sensors, such as electrocardiogram, electroencephalography, blood pressure signals and respiratory signals.
In short, all possible input data may be classified as structured stationary data (e.g. clinical symptoms), structured, sequential data (e.g. electrocardiograms) or unstructured sequential data (e.g. clinical notes in free text).

%Extracting the relevant information from each data type can be done using a combination of three input subnetworks.
%The structured stationary data can be processed by a multi-layer perceptron composed of dense and output blocks.
%Besides, a stack of multiple bidirectional long short-term memory~\cite{schuster_bidirectional_1997} units may process the structured, sequential data.
%Finally, the unstructured sequential data may be processed by bidirectional encoding representations obtained by transformers~\cite{devlin_bert_2019} and multi-layer perceptron.
%All three subnetworks' outputs are concatenated as input to the task-shared subnetwork described in the previous paragraph.

\subsection{Classification of tertiary medical interventions}

Tertiary interventions in spaceflights may be classified as advance life support care, transitional care, ambulatory care, palliative care and emergency care.
The medical intervention classification for any medical situation can be directly mapped from the five predictions carried out by the {\em multitask prediction modeling}.
Hence, a mapping from the five outputs of the multitask prediction modeling described in section~\ref{sec:multitask-prediction-model} and the five types of tertiary medical interventions may be feasible solutions, and it may be obtained by applying a Delphi methodology over a panel of physicians and mission designers~\cite{dalkey_delphi_1969}.

%quizá faltan ejemplos.

\subsection{Optimization of action plans}

Once a compatible diagnosis and the type of tertiary intervention have been assigned, it is time to apply a set of medical actions to restore the crewmembers' health and performance. 
As described in the basic principles, they should be compatible with the Level of Care V for spaceflight missions.
Therefore, the medical actions available in exploration and planetary missions would be an upgraded version of the routine and emergency medical operations included in the ISS Integrated Medical Group Medical Checklist~\cite{mission_operations_directorate_international_2001}.

A complete medical action plan should be tailored to the medical situation, trying to optimize the fundamental mission priorities under the constraints imposed by the restricted clinical equipment and a limited number of medical supplies for treatment.
For this functionality, current solutions based on reinforcement learning~\cite{gottesman_guidelines_2019} may help to optimize the sequence of medical actions by continuously checking their results for the expected clinical outcome of the patient and the mission fundamental priorities.

\smallskip

Emergency events are situations where the crew may more need onboard clinical decision support. Nevertheless, it cannot be isolated from the full tertiary interventions deployed for the medical prevention strategy.
Of most interest is to evaluate if multitask prediction models developed for medical emergencies on Earth~\cite{ferri_deep_2020} can be transferred to onboard decision making with high rates of accuracy.
Doing that, the development of clinical decision support systems to help onboard medical interventions may take advantage of massive biomedical data analysis performed on Earth. 

\section{Space adaptive learning}

As described above, deep learning~\cite{lecun_deep_2015} is the current technology with more success for mimicking human decision making~\cite{silver_mastering_2016} from complex types of data, such as involving high dimensional and multimodal data~\cite{russakovsky_imagenet_2015}, sequences~\cite{hinton_deep_2012} and unstructured data~\cite{hirschberg_advances_2015}.
Given the modular architecture proposed in~\ref{sec:multitask-prediction-model}, we suggest performing an independent learning process of each task-independent subnetwork by the Adam stochastic optimization algorithm~\cite{kingma_adam_2017} with a weigh decay term to promote regularization~\cite{krogh_simple_1991} followed by their ensemble as loosely coupled models~\cite{chen_deep_2015} may result in a feasible static solution for developing the first prototypes of the predictive models.
Moreover, as a basic part of the design, we propose following the methodology proposed by Kohavi~\cite{kohavi_study_1995} to evaluate the model performance and tune hyperparameters without biases maximizing available re-use data.

\medskip

Nevertheless, the main potential limitation to design prediction models for medical decision-making during exploration missions is the continuous medical dataset shift~\cite{moreno-torres_unifying_2012} of the onboard cases produced by the physiological disturbance.
Dataset shift was first described in~\cite{press_dataset_nodate} and defined by Moreno-Torres {\em et al.}~\cite{moreno-torres_unifying_2012} as the situation in which the training and test data follow joint distributions that are different.
Dataset shift occurs when the data experience a phenomenon that leads to a change in the distribution of a single feature, a combination of features or the output boundaries.

In medical prediction problems, where the output (e.g. the disease) causally determines the values of the features (e.g. symptoms), there are two types of dataset shift that may appear independently or at the same time.
First, the prior probability shift refers to changes in the distribution of the output variable. In space medicine, it is observed how the prevalence of some medical conditions increases due to the crewmembers' specific environment and activities. Then, the frequencies of arrhythmia, headache, dermatitis, respiratory infection, and renal stone formation, among other medical events, are increased in space compared to Earth.
We can expect that the probability shift of medical conditions in space will continuously change given the cumulative effect of space influence in the physiological disturbance and the non-stationary environment of exploration missions.
Second, concept drift (a.k.a. concept shift) happens when the representation of the inputs conditioned to the outputs of a predictive model changes in test cases with regard to training cases.
In medical applications, this may happen when the observation of symptoms manifesting diseases changes during operations with respect to the data distributions learned during the prediction model's design.
Given the effect of microgravity, radiation and isolation in human bodies, we can expect a major concept drift in the biomedical data generated in long-term space missions.

\medskip

Hence, designing an effective data-driven decision support system for space exploration missions will require a continual update of prediction models to the latest registered data.
Several alternative approaches can solve that:
a) perform re-training using the complete historical dataset,
b) perform continual learning of models' parameters, including periodically the new test cases~\cite{parisi_continual_2019}, %quiza añadir incremental learning de Salva. %quizá mas informacion sobre parisi: continual learning
c) select the most robust models adapted to imprecise environments~\cite{provost_robust_2001}.
Given that currently there are no registries of biomedical data from humans in space exploration missions, our choice is the continual learning approach that avoids continuous access to historical multisource data, allows using data from ISS and Earth at present and produces prediction models adapted by new observed cases conditioned by the disturbance physiology effects of space.

Besides, a careful monitor of biases affecting data representation should be carried out to ensure the quality and performance of updated models~\cite{rajkomar_machine_2019}. 
The methodology developed in~\cite{saez_kinematics_2018} based on non-parametric statistical manifolds may be useful to calculate the dynamics of temporal variability of biomedical data from crewmembers, including continuous temporal trends, seasonal behaviors and abrupt changes produced by dataset shift effects.

\smallskip

The most challenging step in the design of the MEDEA system is, indeed, adaptive space learning. Added to the lack of data from deep space, continual learning is nowadays an open topic still to be solved for terrestrial scenarios.
Nevertheless, space exploration missions involve a dynamism difficult to compare with other situations. Hence obtaining successful continual learning of onboard data-driven clinical decision making is the most salience hypothesis of the MEDEA concept design.
With our approach, we expect that the {\em evidence of tomorrow will help us further develop and build smart medical systems to address those yet undiscovered challenges of long-duration, long-distance spaceflight~\cite{doarn_health_2019}}.

\section{Semantic interoperability}

%This module is in charge of exchanging unambiguous information with the computer systems to maintain the crewmembers' health.
Space agencies have addressed data exchange in multiple vendors environments by definition of interoperability protocols, such as STEP-SPE. %quizá falta especificar que se ha hecho en industria aerospacial en estandares y estandares de datos medicos
Whereas space agencies put the focus on the exchange of information among space environment analysis tools~\cite{tec-t_gstp_2017},
medical informatics has focused on deploying semantic interoperability in healthcare organizations, which goes some steps further in the exchange of information among heterogeneous systems.
Specifically, semantic interoperability is defined as the transmission of data along with the required knowledge to understand it by sharing clinical concepts described in a reference model using a binding medical terminology shared vocabulary~\cite{tomas_detailed_2016}.
This may allow sending information to buses of data without assuming that every receptor needs to know in advance its semantic.

The standard Fast Healthcare Interoperability Resources (FHIR) was designed for exchanging electronic health records by the Health Level Seven International organization, and the American Medical Informatics Association supports it.
The idea is to organize entities, such as patients, observations, measurements and diagnoses, as FHIR resources specified by profiles (clinical concepts) with U.S. Core Data for Interoperability elements written in the SNOMED Clinical Terms terminology. 
%Specifically, raw data, such as signal monitoring from biosensors, may be received from the onboard health care system,
Adopting of semantic interoperability standards, such as the Fast Healthcare Interoperability Resources (FHIR), to exchange medical data between a clinical decision support system and the onboard healthcare systems and medical diagnosis systems
may ensure the correct interpretation of the large heterogeneous data from crewmembers.
In that way, the astronauts' medical records will be consistent with the international standards, followed by the main providers of medical information technologies.

\smallskip

Although FHIR may provide full semantic interoperability with health information systems, given that current onboard computer systems do not follow interoperable standards yet or they are based on industrial standards from aeronautics, the semantic operability module may also include connectors to the specific onboard systems with adapters to their data structures.

% Crew Health Care System (CHeCS). Their task is to fulfil the overall mission of crew health care} by reinforcing each of the three levels of an astronaut's care. \vbs{The tertiary level of this SLDS hierarchy, the Health Maintenance System (HMS) and included assisting with the final implementations of the Medical Kit Redesign M(KR) and creating and updating various databases used within HMS and CHeCS}. The EMTP (emergency medical treatment) is for use under emergency situations and is used to sustain life. This kit is comparable to the ALSP as it included items such as the AMBU bag, certain medications, and intraosseous (IO) infusion devices. 

%Interoperability for Space Environment Analysis Tools
%https://www.esa.int/Enabling_Support/Space_Engineering_Technology/Shaping_the_Future/Interoperability_for_Space_Environment_Analysis_Tools
%The data needed for the analyses are manifold and tool dependent. A few years ago, the need to define a project data exchange and storage protocol between Space Environment Analysis Tools was identified to improve the consistency of space environments related analysis, permitting projects data tracing and sharing. Following the thermal domain approach, a STEP-based solution was chosen, and a first definition of the STEP-SPE protocol was carried out in a previous TRP study.

\section{Ethical \& legal functional support}

%conclusion: hipotesis de como funcionara
An ethical and legal-based MEDEA design may ensure the practical implementation of ethics responsibilities for spaceflights and ethical standards for artificial intelligence as described by items \#9 and \#10 of the basic principles in section~\ref{sec:principles}.
A careful validation of its performance through computational simulations, test beds, and analogs may guarantee the MEDEA system's correct functionality for human well-being.

\section{Discussion}

\subsection{Medical situations where MEDEA may help}

%%%emergency
The concept design presented here is willing to give real-time decision support for continuously changing medical situations during long-duration spaceflight, with special attention to medical emergencies.
As we have seen in Table~\ref{tab:romerosisisi2}, medical problems, such as cardiovascular accidents, injuries, traumatisms, neurological problems, and anaphylaxis, may arise suddenly in healthy monitored crewmembers with serious consequences to the life of the patients if medical interventions are not activated immediately.
Although the diagnosis may be a simple task for some situations, for others, such as anginas, myocardial infarctions, sudden cardiac arrests, neurogenic and hypovolemic shocks, among others, a differential diagnosis would be required. For other cases, further secondary diagnoses or prognosis would be beneficial.
Moreover, when the medical condition is not a result of an accident, its correct management may require understanding its cause. For example, idiopathic illnesses may be associated with the changing environment in the space, subclinical diseases and genetic predispositions.
Finally, regardless of the diagnosis, an optimal treatment plan would be essential to initiate the patient's recovery, taking into account the patient's condition and the crew's safety throughout the mission.

\medskip

Other authors have studied medical situations in space missions, obtaining similar to us.
Kuypers {\em et al.} in~\cite{kuypers_emergency_2013} include two lists of health concerns for specific space conditions and medical emergencies, respectively.
The first list identifies idiopathic and subclinical diseases such as cancer, cataracts, immunologic changes, decreased red blood cell mass, bone and mineral loss, muscle atrophy/loss of strength, vestibular/sensorimotor changes, cardiovascular changes, hyperopic vision shifts, mental health problems, bacterial growth, water and air contamination or degradation, and other deficiencies.
Whereas, the second list includes emergency conditions due to occupational and environmental factors such as wounds, burns, contusions, sprains, fractures, cardiac dysrhythmias, orthostatic intolerance, pneumonitis, persistent latent viral reactivation, anaphylactic reactions, dermatologic cellulitis, dermatitis, space motion sickness, gastroenteritis, constipation, renal stones, urinary tract infections, acute urinary, retention, crown fracture, dental infections, abscess, corneal abrasion, corneal infection, foreign bodies, depression, anxiety, exposure to toxins, acute radiation illness and decompression sickness.
Besides, Stewart {\em et al.}~\cite{stewart_emergency_2007} identified traumatic injury as one of the most relevant emergencies in space exploration given the expected incidence and consequences to the mission.
Additionally, they pointed out the lack of knowledge about the cardiovascular and immunological effects of long-duration spaceflight on wider medical profiles of astronauts.
%% ~\cite{blue_challenges_2019} Fever, infection, and hematopoietic sequelae. Infection remains a concern in the spaceflight environment, particularly as micro-gravity conditions are known to be associated with immunosuppression and increased clinical sequelae risk.27,56,121
%Besides, Stewart {\em et al.} in~\cite{stewart_emergency_2007} and Komoroski and Fleming in~\cite{komorowski_intubation_2015} focused on the medical emergency procedures of the critically ill and injured in extreme conditions and environments.

% decir que en todas estas situaciones, nuestro sistema operarará

\subsection{Potential benefits of MEDEA compared to other approaches}

Facing the impossibility in the current time of performing feasible experiments to compare approaches for managing medical emergencies during human space exploration, we have directly compared the potential benefits of the proposed software suite to those obtained by a human medical approach and the use of clinical decision support systems designed for terrestrial use operating remotely from Earth or on board.
As it is shown in Table~\ref{tab:comparison}, an onboard clinical decision support system does not substitute human medical professional for performing physical interventions on the injured crewmembers. Nevertheless, several benefits may be expected from the functionalities of an adaptive clinical decision support system.
Medical decision making without delay can be carried out by a professional on board for saving lives in case of life-threatening health situations, but in case no medical professional is present or fit to assist, the software suite may give advice and guide without delay crewmembers that do not have such specific skills.
Moreover, given the physiological disturbances and changing conditions to which crewmembers are subjected, a continuous update of medical knowledge is required to perform an accurate evaluation of health situations. In this way, static decision models designed for terrestrial conditions may be obsolete soon when deployed on board of the spaceflight, so decision support systems continuously adapted to dataset shifts may be required to maintain good levels of performance in decision advice.
Other relevant benefits of using decision support systems adapted to space exploration missions may come by the intrinsic compliance of decision-making with the space missions' formal conditions, such as the mission fundamental priorities and ethics responsibilities. Moreover, long-term scheduling may benefit from using decision support systems optimized to onboard medical supplies and equipment.
Finally, another benefit of the proposed approach is the compliance of ethical standards for artificial intelligence, such as good practices for data protection, fairness, interpretability and traceability.

\begin{sidewaystable}
\caption{Benefits of MEDEA compared to solutions based on human medical professionals, terrestrial CDSS on Earth and terrestrial CDSS on board of the spaceflight. \label{tab:comparison}}
\footnotesize
\begin{tabular}{p{5cm}|p{3cm}p{3cm}p{3cm}p{3cm}}
	\hline
	Features & Human medical professionals & Terrestrial CDSS on board & Terrestrial CDSS on Earth& MEDEA\\
	\hline
	Perform physical interventions on the injured & Yes, if trained or assisted & No, electronic clinical guidelines may assist & No, no feasible given round-trip delay & No, electronic clinical guidelines may assist\\
	\hline
	Feasible to decide about medical emergencies on board & Yes, if one active physician on board & No specific to spaceflights & No specific to spaceflights & Specific to spaceflights \\
	\hline
	Autonomous decision & Yes, if one active physician on board & Yes, computation on board & No, computation on Earth & Yes, computation on board\\
	\hline
	Patient-specific advice & Yes, based on expert reasoning & No, designed under Earth conditions  & No, designed under Earth conditions & Yes, continuously adapted to spaceflight conditions\\
	\hline
	Response time & If physician on board, then immediate, else round-trip communication to mission control  & Immediate computation on board & Round-trip delay to mission control & Real-time computation on board\\
	\hline
	Adapted to physiological disturbance and changing conditions & Yes, human reasoning & No, static model from Earth conditions & No, static model from Earth conditions & Yes, adaptive learning \\
	\hline
	Optimized to mission fundamental priorities & Yes, human reasoning & No & No & Yes, Artificial Intelligence \\  
	\hline
	Optimized to medical supplies and equipment on board & Human reasoning & No & No & Yes\\
	\hline
	Applying health standards for levels of care V & Yes, mission protocols & No & No & Yes, Artificial Intelligence\\
	\hline
	Implement ethics responsibilities for spaceflights & Yes, mission protocols & No & No & Yes, Artificial Intelligence\\
	\hline
	Applying ethical standards for artificial intelligence & No applicable & Maybe  & Maybe & Yes, data protection, fairness, interpretability and traceability\\
	\hline
\end{tabular}
\end{sidewaystable}

\subsection{Potential impact} %\medskip

%%% Impact of the proposal

%impact 1: tertiary medical interventions for space exploration missions

In 2010 and 2019, the cost of the first mission to Mars was estimated at 6 Billion USD and 10 Billion USD, respectively.
Added to the loss of lives, a health problem in the crew may risk the rest of the crewmembers, jeopardize the mission and loss the vehicle and payload.
{\em History has shown that during the exploration of frontiers on Earth, human physiologic maladaptation, illness, and injury have accounted for more failures of expeditions than any single technical or environmental factor~\cite{medicine_safe_2001}}.
Therefore, it is critical to rely on robust fault-tolerant solutions to deploy tertiary medical interventions most autonomously during space exploration missions, where evacuation and synchronous communication to mission control is not a reliable option.

Clinical decision support systems may provide real-time advice tailored to the health problem and optimized the mission's fundamental priorities and compliance with the highest ethical and legal conditions.
The support for a wide set of medical conditions must be deployed to deal with medical emergency events and nominal health issues as well, in light of the expected increase of astronauts' profiles and civilian spaceflight~\cite{stepanek_space_2019}.
Moreover, adaptive learning must guarantee predictive models updated to the cumulative and variable disturbance of space effects in human physiology.

\medskip

%impact 2
The autonomous clinical decision support system developed in MEDEA may be directly transferred to medical applications on Earth.
Specific scenarios may have special requirements similar to space exploration missions. Isolation and extreme conditions may appear on deep-sea exploration, Arctic and Antarctic missions, and isolated communities on desserts and forests.
The global market of clinical decision support systems for general and specialized medicine for 2028 has been estimated at 3.5 Billion USD.
The development of quantitative medicine will go hand-in-hand with this technology's design, so robustness and adaptation required for operating in space will speed up this process.

\medskip

%impact 3: a step forward to the future healthcare
Sooner or later, humans will expand along with the Solar System, making spaceflights among planets, asteroids and space stations the next reality of human beings.
Developing adaptive clinical decision support systems will be a keystone for delivering medical care on board, opening the career of healthcare in space.
It is impossible to estimate healthcare's economic impact after humans expanse, but we can grossly understand its dimension by the size of the current global healthcare market calculated at 12 Trillion USD.

\medskip

%%% Limitaciones y challenges
Getting CDSS on board a spacecraft is not going to be easy. The first versions of software as MEDEA, will have to be designed simultaneously as the missions to Mars are designed for the launch windows by 2030s.
Therefore, integration with onboard health information systems may represent a challenge on its own.
Moreover, the initial versions of predictive models will not be able to be trained from any real data acquired during previous planetary missions, so we will have to use medical data from Earth, analogs, virtual analogs and ISS for the first versions of the prediction models; relying on the continual adaptive learning during the trip to get relevant knowledge from the physiological conditions that astronauts will experiment on board of the spaceflight.
The principles and concept design described in this study are intended to serve as the basis to implement a complete and qualified clinical decision support system to operate in space exploration missions before the 2030s. 

\section{Conclusion}
In less than two decades, space missions for human exploration of Mars will be designed and launched.
A key element for their success will be providing autonomous healthcare adapted to dynamic space conditions.
For that, onboard healthcare may be fully redesigned, considering that low-latency telemedicine and that prompt evacuation to Earth will not be feasible.
Therefore, autonomy for real-time decision making will be mandatory to solve medical emergency situations and monitor medical status for space induced health conditions.

This study introduced the basic principles and concept design of MEDEA, a clinical decision support system to provide real-time advice to tertiary interventions on board of space exploration missions.
The presented design applies the current Biomedical Data Science and Artificial Intelligence technology to develop clinical decision making to support onboard medical emergencies by fulfilling the fundamental priorities and the level V of healthcare in exploration spaceflights.
The design consists of four dependent modules, being the main one responsible for giving direct advice to the crew using learning multitask network to predict the medical event's character, a classifier of the tertiary medical intervention and an optimizer of the medical action plan.
A continual deep learning module will solve the prediction model's adaptation to the changing physiology on space, and both modules will be integrated with onboard health information systems using semantic interoperability.
Fairness, interpretability and traceability of decision making may ensure best practices and trust during the full operational time of MEDEA.

The clinical decision support system implementing the MEDEA concept design is expected to give autonomous decision making for the next human missions to Mars. Besides, it will represent a stepstone for the future of quantitative medicine on Earth and a potential healthcare device to expand humankind throughout the Solar System.

\section*{Acknowledgements}

The author thanks Pablo Ferri, Vicent Blanes, Pere Blay, Alberto Conejero, Alberto González, Elies Fuster and Carlos Saez for their comments to the concept design described in this study. Of special value were the clinical discussions with Antonio Felix, Felipe Mesquida, Ramon Puchades and Nekane Romero.

\section*{Funding sources} 

This research did not receive any specific grant from funding agencies in the public, commercial, or not-for-profit sectors.

\section*{Competing interests statement}

The author has no competing interests to declare.

\bibliographystyle{elsarticle-num} 
\bibliography{medea}

\end{document}